# Skyrmion Lattice Phases in Thin Film Multilayer.

Jakub Zázvorka[1,2, +], Florian Dittrich[1, +], Yuqing Ge[1], Nico Kerber[1,3], Klaus Raab[1], Thomas Winkler[1], Kai Litzius[1,3,4], Martin Veis[2], Peter Virnau[1,3,*], Mathias Kläui[1,3,*].

1. Institut für Physik, Johannes Gutenberg-Universität Mainz, Mainz, Germany.
2. Charles University, Faculty of Mathematics and Physics, Institute of Physics, Ke Karlovu 5, CZ-121 16, Prague 2, Czech Republic.
3. Graduate School of Excellence Materials Science in Mainz, Mainz, Germany.
4. Max Planck Institute for Intelligent Systems, Stuttgart, Germany.

[+]*These authors contributed equally.*

[*] Corresponding authors: Klaeui@uni-mainz.de, Virnau@uni-mainz.de

**Abstract**

Phases of matter are ubiquitous with everyday examples including solids and liquids. In reduced dimensions, particular phases, such as the two-dimensional (2D) hexatic phase and corresponding phase transitions occur. A particularly exciting example of 2D ordered systems are skyrmion lattices, where in contrast to previously studied 2D colloid systems, the skyrmion size and density can be tuned by temperature and magnetic field. This allows us to drive the system from a liquid phase to a hexatic phase as deduced from the analysis of the hexagonal order. Using coarse-grained molecular dynamics simulations of soft disks, we determine the skyrmion interaction potentials and we find that the simulations are able to reproduce the full two-dimensional phase behavior. This shows that not only the static behavior of skyrmions is qualitatively well described in terms of a simple two-dimensional model system but skyrmion lattices are versatile and tunable two-dimensional model systems that allow for studying phases and phase transitions in reduced dimensions.

**Introduction**

Magnetic skyrmions, topologically stabilized whirls of magnetization, are in the focus of the scientific community due to their attractive properties for possible novel application devices[1–3]. Using spin-transfer torque and spin-orbit torque[4–7], skyrmions can be moved with high speeds at low current densities and can even be stabilized with no external magnetic field applied[7–9], which makes them potentially useful for memory and computer logic devices[3,10]. In particular for reservoir computing[11], the skyrmion interaction and collective behavior is of key importance. Skyrmion lattices have been found widely in bulk materials with B20 symmetry[12], where the topological structures are stabilized due to bulk Dzyaloshinskii-Moriya-interaction (DMI)[4,5,13]. However, in bulk systems the skyrmions are mostly not 2D like, as the "skyrmion tube" length can easily exceed the skyrmion diameter. In thin film systems, skyrmions down to sub-nm thickness and diameters in the range of micrometers are stabilized, making them prime candidates for perfectly 2D systems. While skyrmion lattices have been studied theoretically in such systems[14,15], only recently first experimental reports of thin film lattices have been reported, albeit with systems where the relatively large (~100 nm) film thickness is similar to the lateral skyrmion size making these systems not necessarily 2D[13,16]. Thus to experimentally probe the rich phase behavior of two-dimensional systems[17–19] akin to colloids in the past[20–22], 2D skyrmion lattices occurring in ultra-thin film stacks might be an ideal model system[23]. As the nature of phase transitions in two-dimensional systems of hard and soft disks has been a grand challenge in Statistical Physics, which has recently been numerically treated[17], there is a clear need for experimental 2D systems to experimentally probe the phase behavior. This calls for studying 2D skyrmion lattices and analyze their phase behavior with numerical simulations based on coarse-grained models from Statistical Mechanics to identify possibly unique 2D properties as well as gauge the suitability of these systems to study the exciting 2D phase behaviors.

Thus, in this work, we use a sub-nm thick CoFeB-based multilayer system to study the emergence of skyrmion lattices as well as their response to tuning external parameters such as temperature and field. Since the skyrmion diameter (Fig. 1a) is three orders of magnitudes larger than its thickness, this system could be considered to be inherently 2D. By experimentally ascertaining the phase transitions we demonstrate the 2D nature of the system as well as its suitability as a model system to probe 2D phase behavior.

**Results**

Using Kerr microscopy imaging we investigated a low-pinning multilayer stack Ta(5)/Co$_{20}$Fe$_{60}$B$_{20}$(0.9)/Ta(0.08)/MgO(2)/Ta(5) similar to a material previously characterized in which the skyrmions show thermally activated diffusion at low skyrmion densities[24]. The studied material exhibits perpendicular magnetic anisotropy (PMA). Using out-of-plane magnetic field sweeps, stripe domains and a low density of skyrmions are present in the sample. Upon fixing the out-of-plane field and a subsequent saturation of the sample using an in-plane field in any direction, a high density of skyrmions is nucleated in the sample when the in-plane field is reduced back to zero abruptly. The density and the mean radius of the skyrmions is controlled by the values of the out-of-plane magnetic field applied and the temperature. For details on the MOKE hysteresis loops and skyrmion lattice nucleation, see Supplementary Information. By varying the out-of-plane field, the size and as a result also the skyrmion lattice density and ordering is tuned, which is a unique handle compared to previously used systems, such as colloids with fixed sizes.

Variations in temperatures are found to tune the amount of thermally activated motion[24] but also the average skyrmion radius, as well as lattice density due to the changing magnetic properties. To evaluate phases in 2D systems such as the skyrmion lattice phases (Fig. 1a), we employ two quantifiers:

The local orientational order parameter[17] (Fig. 1b)

$$\psi_6(k) = \frac{1}{n_k}\sum_{l=1}^{n_k} e^{i6\theta_{kl}} \qquad \text{Eq. (1)}$$

and the pair correlation function (Fig. 1c)

$$g(r) = \frac{1}{4\pi r^2}\frac{1}{N\rho}\sum_{k=1}^{N}\sum_{k \neq l}^{N}\langle\delta(r - |\mathbf{r}_k - \mathbf{r}_l|)\rangle \qquad \text{Eq. (2)}$$

The local orientational order parameter is a standard measure to quantify the emergence of local hexagonal order[17]. $\theta_{kl}$ describes the angle of the connecting line between a skyrmion and one of its $n_k$ nearest neighbors with respect to a fixed axis. For a perfect triangular lattice, the contribution of all six neighbors yields $|\psi_6|=1$. The one-dimensional pair correlation function g(r) (Eq.2) contains basic information such as typical nearest and next-nearest neighbor distances and the general structure of a gas, liquid or crystal. We use this parameter in our modelling approach to reproduce the basic structure of the system while keeping the fitting procedure manageable. Eq. (2) is, however, not suited to visualize the emergence of hexagonal order like the 2D-pair correlation function used[17]. We provide more information on these quantifiers in the Materials and Methods section.

To study the evolution of the phases of the system, we take a video using the Kerr microscope after an in-plane magnetic field is switched off. The observed skyrmions are tracked, their positions are evaluated, and are quantifiers calculated for each frame in the video. Calculation of the correlation functions and individual skyrmion position evaluation is described in the Materials and Methods section. The local orientational order parameter is calculated for every skyrmion in one frame except for those on the border of the frame. Note that in this context the expression "order parameter" refers to a parameter which quantifies the local orientational order of a system and is not to be understood in the classical sense as a parameter which characterizes second order phase transitions. To obtain a quick indication of the state and the phase of the system, we introduce a heuristic parameter $\langle|\psi_6|\rangle$, which averages the absolute value of $\psi_6$ over all skyrmions for which $\psi_6$ was computed. From a soft disk system[17] we find that $\langle|\psi_6|\rangle \approx 0.69$ indicates the transition from a liquid phase (below) to an hexatic phase (above) as shown in Supplementary Information.

Figure 1b shows $\langle|\psi_6|\rangle$ of the skyrmion lattice at fixed out-of-plane field and sample temperature as a function of time after the initial lattice nucleation. As visible, the angular ordering as well as translational ordering as quantified by the pair correlation function (Figure 1c) is not constant instantly after switching off the magnetic in-plane field. Immediately after switching off this field, the skyrmions are nucleated on a timescale that is below the time resolution of the measurement setup (ms). This is then followed by a stabilization phase in the range of seconds to tens of seconds. While the initial ordering occurs rather quickly in all cases, $\langle|\psi_6|\rangle$ is still increasing slightly over the course of our measurement (60s) consistent with the expected prolonged equilibration times associated with the emergence of hexagonal order. We refer to the last 30s of the $\langle|\psi_6|\rangle(t)$ as the "semi-steady" state, which is a sufficiently long period to robustly measure quantifiers. The initial 4s after switching off the in-plane field where the highest slope $\langle|\psi_6|\rangle(t)$ is found and where the nucleated skyrmions form a lattice is referred to as the "nucleation" period. The "relaxation" period covers the remaining part of the time evolution.

While as shown in Fig. 1b, at 338 K the system orders with $\langle|\psi_6|\rangle > 0.69$ (indicating possibly a hexatic phase, see further below for a detailed discussion), at 330 K $\langle|\psi_6|\rangle$ only goes up to the value of 0.55, consistent with the formation of a more disordered dense liquid phase. Likewise, the pair correlation function also changes in the course of equilibration (see Fig. 1c). Fluctuations are related to the thermally activated movement of the skyrmions that occurs in the lattice. We observe that skyrmions repel each other and we do not see any significant skyrmion-skyrmion annihilation thus boding well to study the phase transitions.

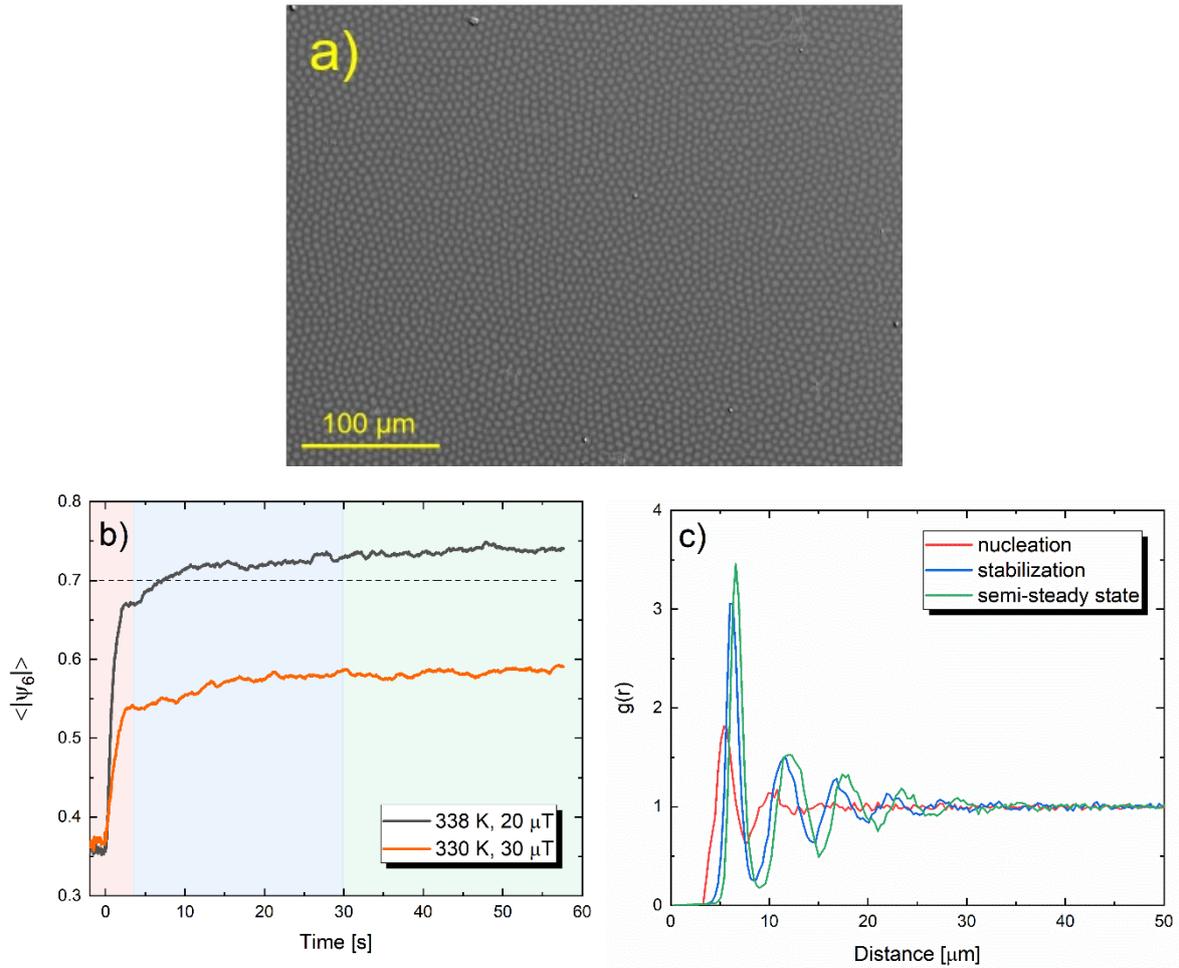

**Figure 1**. Picture of skyrmion lattice and evolution of phase quantifiers. (a): Kerr image of a skyrmion lattice at 335 K with μm sized skyrmions. (b) The evolution of $\langle|\psi_6|\rangle$ averaged over all skyrmions in one frame in dependence of time after nucleation. The red, blue and green backgrounds depict the nucleation, stabilization and a semi-steady skyrmion state, respectively. $\langle|\psi_6|\rangle$ is dependent on the temperature and the applied out-of-plane field. c) Pair correlation function g(r) right after nucleation, in the stabilization phase and in the semi-steady state for temperature 338 K and 20 μT applied out-of-plane field. After switching off the in-plane field and the resulting skyrmion nucleation, the red pair correlation function (in Fig. 1c) emerges and indicates typical nearest and next-nearest neighbor distances. Blue and green curves show the pair correlation function g(r) in the relaxation and semi-steady state of the lattice, respectively. In the stabilization process the correlation function is noisier, whereas in the semi-steady state the function has a finer distribution.

Having established the time evolution of $\langle|\psi_6|\rangle$, we now systematically study of the dependence of the semi-steady state lattice properties on the external parameters, temperature and magnetic to explore the tunability. The average $\langle|\psi_6^t|\rangle$ is obtained from all frames after 30 seconds of equilibration and shown in Figure 2. With reducing temperature, the range of out-of-plane field where skyrmions can be stabilized becomes narrower and a monotonic trend of higher hexagonal order with higher temperature is observed. The highest temperature achievable was limited by the measurement temperature control as well as the spatial resolution since the skyrmion diameter depends on temperature. At too low out-of-plane fields, after in-plane field sweeps, not only skyrmions are stabilized but also elongated "worm" domains are present. These effectively distort the lattice and hinder its higher ordering so that we have focused on parameter combinations where we have only skyrmions. A decreasing tendency of angular order is found at increased out-of-plane field values for every studied temperature. This can result from higher skyrmion-skyrmion distances, where the thermal movement of the magnetic textures is more pronounced and thus hinders the ordering of the lattice. A maximum value of $\langle|\psi_6^t|\rangle$ of around 0.73 is obtained at the highest investigated temperature and 20 µT out-of-plane field when also the highest observable skyrmion density is reached.

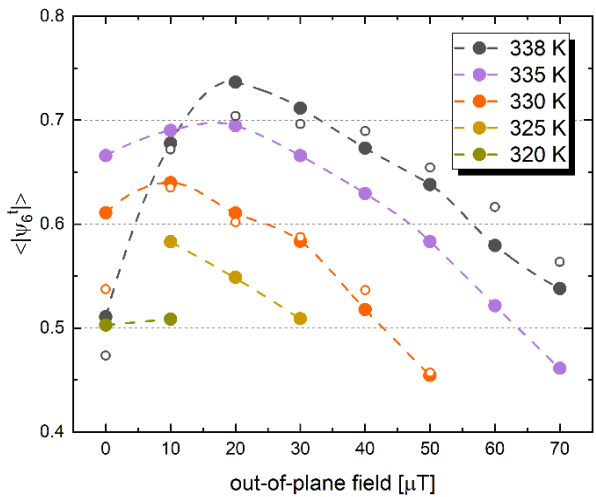

**Figure 2**. Time averaged $\langle|\psi_6^t|\rangle$ for different out-of-plane fields in the temperature range 325 – 338 K. The field and temperature ranges are limited by the stability of a pure skyrmion lattice as well as the minimum size of skyrmions that can be detected. The highest ordering achieved is at 338 K with 20 µT. $\langle|\psi_6^t|\rangle$ was calculated from the skyrmion position in the sample in the semi-steady state part of the skyrmion lattice formation (after 30 seconds since the skyrmion nucleation). Empty circles are simulation results corresponding to T = 338 K and T = 330 K and were determined after $10^6$ simulation time steps. Dashed lines serve as guidelines between points only.

As the onset of the hexatic (or even solid) phase is directly visible in the spatially resolved map of the local orientational order parameter, we study this at the maximum value of $\langle|\psi_6^t|\rangle$ (338 K, 20 µT): Figure 3a shows the hexatic skyrmion domains with coincident orientation of $\psi_6$ as measured by the angle $\theta$ (essentially the Euler angle of $\psi_6$ divided by 6 as explained in Materials and Methods). The average domain size is of the order of 50 µm, corresponding to roughly 100 skyrmions. In particular we see a homogenous distribution of $|\psi_6|$ in Fig. 3c.

For comparison, we also show the corresponding liquid phase results for T = 330 K and B = 40 µT in Figures 3b and d). Note that skyrmions are much larger under these conditions and domains of similar orientation are of the order of 10 particles or less. In this liquid phase there is no homogeneous distribution of $|\psi_6|$ as shown by the irregular colors in Figure 3d.

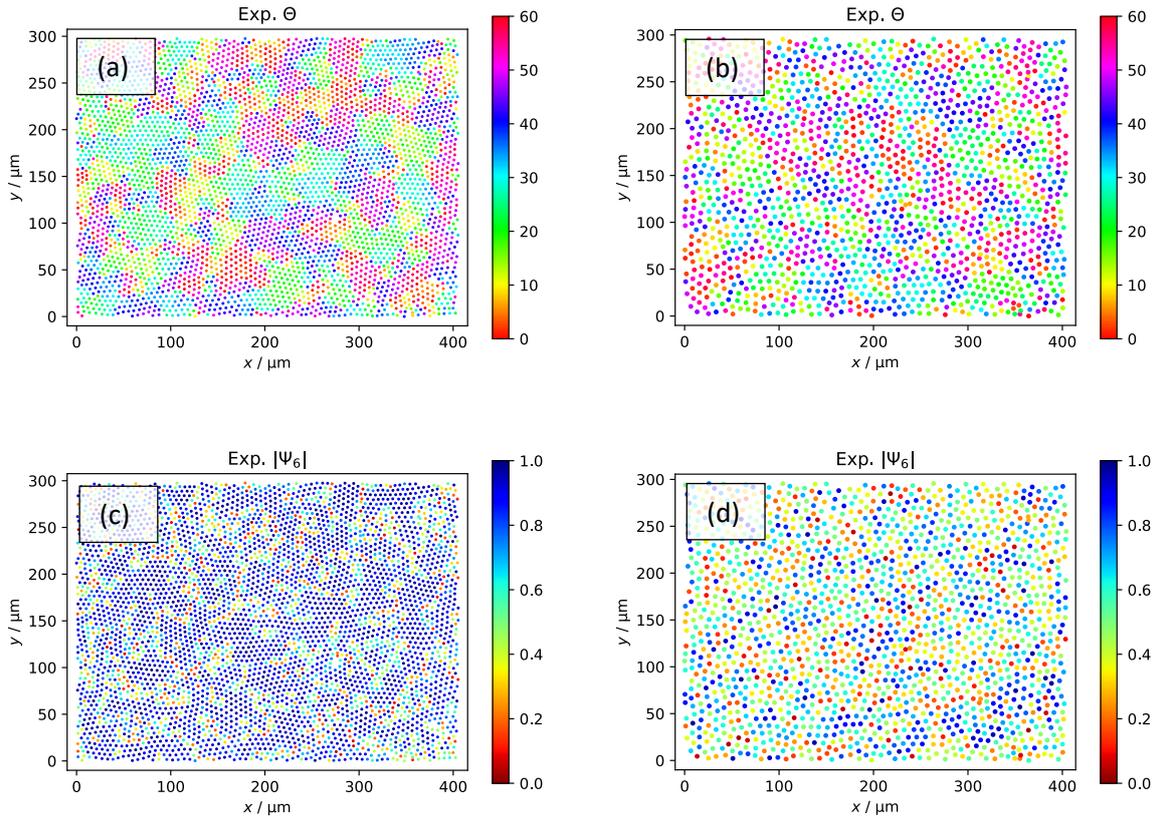

**Figure 3**. Spatial distribution of the local orientational order parameter $\psi_6$ of individual skyrmions. (a) and (c) were evaluated at 338 K and an out-of-plane field value of 20 µT. This represents the state with the highest value of $\langle|\psi_6^t|\rangle$ in Fig.2. (a) visualizes the orientation of $\psi_6$, i.e. the orientation angle $\theta$, while (c) visualizes the absolute value of $\psi_6$. (b) and (d) are corresponding figures for 330 K and 40 µT.

To understand our results and draw robust conclusions about the phases and the two-dimensional nature of the studied system, we support the experimental results with numerical simulations using a model of soft particles which interact with each other via a repulsive power-law potential $r^{-n}$. This choice is purely empirical but allows us to describe the strong short range repulsive interaction studied previously[25]. At the same time, the chosen potential benefits from the availability of exact phase diagrams for a wide range of $n$[17]. For $n \geq 6$ (which includes the hard disk scenario) the transition from the liquid to the hexatic phase was shown to be of first order followed by a continuous transition to the solid phase[17]. For smaller values of *n*, the transition from the liquid to the hexatic phase becomes continuous and of the Kosterlitz-Thouless-Halperin-Nelson-Young (KTHNY) type[26].

In the following we want to ascertain to which extent skyrmion lattices can be used as generic model systems to explore the phase behavior of two-dimensional systems akin to colloids[20–22]. At the same time, we want to gauge if a coarse phenomenological model from Statistical Physics is actually able to describe the bulk macroscopic behavior of skyrmions accurately. Building upon expansive numerical work[17], which has determined the phase behavior of soft disks with great accuracy, Molecular Dynamics simulations of this model were performed and mapped onto our skyrmion system. Parameters were adjusted to match the pair correlation function of skyrmions for a given density. Note that fixing *n* (to e.g. 6 to represent interactions between dipoles) will generally lead to a worse agreement with the experimental g(r). For a detailed discussion of the mapping procedure, see Methods and Supplementary Information.

Using this ansatz, we have reproduced the experimentally observed behavior of $\langle|\psi_6^t|\rangle$ for T = 338 K and T = 330 K (Figure 2). Qualitative agreement between simulations and experiments is found. However, one should note that $\langle|\psi_6|\rangle$ is very sensitive to the details of the mapping (e.g. if all details or only parts of g(r) are used). Another caveat for both simulations and experiments at T = 338 K is the time after which $\langle|\psi_6|\rangle$ is measured as it increases during the course of equilibration. Nevertheless, considering that our mapping is purely based on basic structural information (namely density and the one-dimensional g(r)), the qualitative agreement shows that static properties of skyrmion interactions can indeed be captured by a coarse-grained phenomenological model.

A more quantitative approach relies on the decay of the spatial correlation function $G_6$ [20,21] which can also be used to distinguish phases in two-dimensional systems:

$$G_6(r) = \frac{1}{n_r} \sum_{k,l} \psi_6(r_k)\psi_6^*(r_l) \qquad \text{Eq. (3)}$$

In figure 4, we compare the decay of G$_6$ from experiment at T = 338 K and B = 20 µT and simulation. While this correlation function decays exponentially in the liquid phase, quasi-long range orientational order is expected to emerge in the hexatic phase[26]. Depending on the equilibration time after which the correlation function is measured in the simulation, the envelope of G$_6$ increases towards an algebraic decay. We also observe that the experimental data (black dashed curve) is still decaying exponentially and is likely not fully equilibrated, yet.

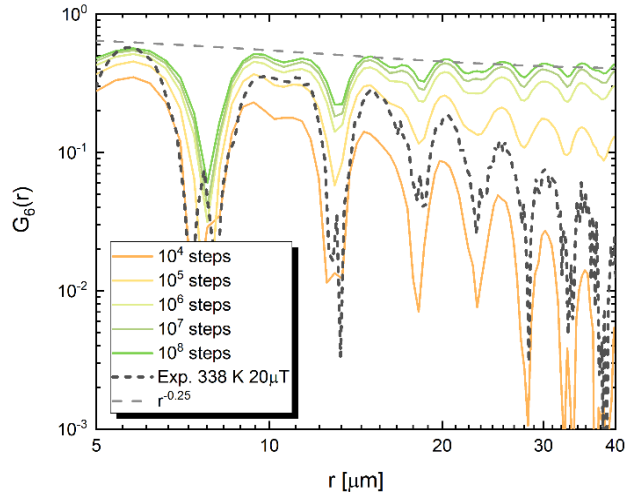

**Figure 4**. Decay of the spatial correlation function $G_6$ for the experimental system (338 K, 20 µT, averaged over frames 300-960) and matching simulations after different runtimes (single snapshot of a quadratic simulation box containing 40000 particles).

The effect of equilibration can also be seen in simulation snapshots. While after $10^4$ equilibration steps the distribution of θ in Figure 5a (as well as the decay of $G_6$) is similar to the corresponding experimental plot (Figure 3a), the domains of similar orientation continue to grow as indicated by a snapshot taken after $10^8$ equilibration steps (Figure 3b).

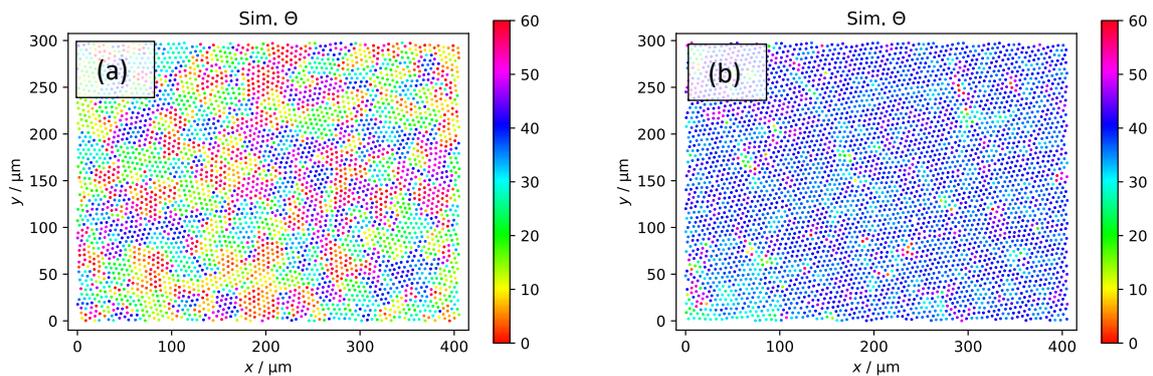

**Figure 5**. Spatial distribution of $\theta$ in a simulation corresponding to a sample temperature of 338 K and an out-of-plane field of 20 µT after $10^4$ (a) and $10^8$ (b) equilibration steps. Only a small part of the simulation box is shown to make plots comparable to Fig.3a.

**Discussion and Conclusions**

Based on our numerical simulations, we conclude that the observation of multiple domains in the experiment (Fig. 3) is likely the result of an incomplete equilibration process as equilibration times are notoriously large in an emergent hexatic (or solid) phase. This is corroborated by the observation that the sizes of the experimental domains continue to grow up to the maximum time which can be measured (that is limited by the setup stability). Additionally, we occasionally see structural defects that pin certain skyrmions that are thus not ordered locally and remain unordered potentially leading to artificial domain wall pinning.

In conclusion, we have analyzed the phases of skyrmion lattices to identify the reduced dimensionality of this µm sized but sub-nm thick systems. We have shown that by using the pair correlation function and local orientational order parameter we can characterize the skyrmion lattice system, which allows us to investigate two-dimensional phase transitions. Temperature and out-of-plane field impact density and mean skyrmion-skyrmion distance and translate to different nucleation dynamics and hexagonal ordering of the observed lattice. We find that the hexagonal order increases with higher temperature and field values in the range of 10-30 µT. Above 338 K, the skyrmion lattice cannot be resolved with the optical microscope setup. For the majority of the selected parameters, we observe behavior consistent with a two-dimensional, dense liquid. However, we also find that for selected conditions, our system is in an emergent hexatic (or even solid) phase showing its 2D nature. As expected for the hexatic phase, we find that the equilibration in this phase is very slow. By comparison with theory, we were able to reproduce qualitatively the experimentally observed phase behavior using computer simulations with a simple phenomenological model based on soft disks by matching density and the one-dimensional pair correlation function. We thus demonstrate that static behavior of skyrmion ensembles may be described by a simple 2D model system highlighting that our skyrmion lattices can indeed be used as 2D model systems with major advantages in terms of tunability and speed compared to conventionally used 2D model systems such as colloids.

**Materials and Methods**

Sample parameters

The sample was prepared using magnetron sputtering in a Singulus Rotaris sputtering tool. The base pressure during the growth process was less than $3\times10^{-8}$ mbar. The composition of the single stack was Ta(5)/Co$_{20}$Fe$_{60}$B$_{20}$(0.9)/Ta(0.08)/MgO(2)/Ta(5), with the thickness of individual layers given in nanometers in parentheses. With the used deposition system, we can tune the thickness of the individual layers in the stack in a controlled manner with high accuracy. The stack is similar to the one reported on previously. The sample was characterized using the magneto-optical Kerr effect (MOKE) measurement. Single skyrmions can be stabilized in the ferromagnetic layer using out-of-plane sweeping. Worm-like domains and skyrmions exhibit thermally activated diffusion. The hysteresis loop in applied out-of-plane field shows an hour-glass shape, typical for material with the presence of skyrmions (see SI). With higher temperature the hysteresis loop is tilted towards larger applied fields. This indicates a change of the anisotropy of the material with temperature. The lowest investigated temperature is determined by the ability to stabilize the skyrmion lattice. Below 325 K, only stripe domains were nucleated by the saturation of the sample with an in-plane field. The highest achievable temperature for the lattice investigation was determined by the resolution of the microscope and the thermally activated motion of skyrmions. Above 338 K, the size of the skyrmions was comparable to the resolution of the Kerr microscope. The skyrmions movement was also more rapid. Above this temperature, no reliable skyrmion tracking in this material system could be performed.

Measurement setup

A commercial Evico GmbH MOKE microscope was used. The in-plane field coil was supplied from the microscope manufacturer. Highest achievable in-plane field was 300 mT. The coil for the out-of-plane field application was custom built at the University of Mainz. The coil was designed to have negligible coercivity and to be able to supply the sample with very small controlled fields in orders of µT. A stage with two QC-32-0.6.1.2 Peltier elements was used for the temperature change of the sample in the range of 280 – 350 K. The temperature was externally controlled by measurement of resistivity of a Pt100 resistor, which was placed next to the multilayer sample. The stability of the set temperature was measured to be within 0.3 K. The frame rate of the microscope camera was 16 frames per second; therefore, the time resolution of the microscope measurement was 62.5 ms.

Skyrmion tracking

Skyrmion lattices are visualized using a Magneto-Optical Kerr-Microscope. In the pictures the out-of-plane magnetization is represented by a grey scale, so that the skyrmions appear as light blobs on a dark(er) background. We consecutively analyze videos recorded that way using the Trackpy [*1] package. In a first stage it locates the skyrmions by detecting Gaussian-like blobs in the grey scale movies.

Several parameters are set to optimize the recognition for reliable results. Most importantly the mask-parameter sets a rough estimate for the pixel-diameter of the features to be found. During our evaluation it is set slightly above the average skyrmion diameter determined by simple binarization of the frame. The separation-parameter enforces a minimum separation between the recognized features, this way over-recognition in defective areas is prevented. A safe value for the recognition is several pixels lower than the average skyrmion distance. The percentile-parameter depends on the contrast of the video and indicates to which extend the features are expected to be brighter than the surrounding area. The noise-parameter is a measure for the "sharpness" of the features to be detected and can vary between measurement videos with different external parameters.

Most of the skyrmion diameters are in a range from 7 to 13 pixels. For example, at the temperature of 338 K and the out-of-plane field of 20 µT, we set the mask to 9, the separation to 4, and the noise to 0.15.

Quantifiers for phase transitions in two-dimensional systems

The *pair correlation function* (PCF) (Eq. 2) determines the probability of finding two skyrmions at a distance **r** from each other. The position of the first peak assesses the mean nearest neighbor distance and deep in the solid phase characteristic sharp peaks resulting from the underlying lattice appear.

It is, however, impossible to distinguish g(r) of liquid, hexatic and solid phases close to the phase transition and other identifiers need to be considered. Since the disk-shaped skyrmions form out hexagonal order as the bulk density increases, one can resort to the *local orientational order parameter* $\psi_6$ [26] (Eq. 1). This complex parameter measures deviations from hexagonal order. The absolute value $|\psi_6|$ = 1 for a perfect triangular lattice and decreases to 0 with increasing disorder. The cut-off distance to find the nearest neighbours was selected to be the position of the first minimum in the corresponding pair correlation function g(r). A strict cut is implemented, so the number of neighbors $n_k$ will usually but not necessarily be 6. In addition to the absolute value of $\psi_6$ one can also extract the local orientation angle of neighboring skyrmions, i.e. the Euler angle of $\psi_6$ divided by 6. Note that the orientation of a hexagonally ordered cluster consisting of the central particle and its six neighbors is essentially determined by the angle between the x-axis and the vector of the central particle and its neighbor in the range of 0 to 60 degrees. The factor of six in the definition of $\psi_6$ projects all vectors between the central particle and its neighbors on top of each other and $\psi_6$ averages over these projections. The orientation angle (ranging from 0 to 60 degrees) is therefore a gauge for the local orientation of the cluster with respect to the x-axis. This parameter is well-suited to visualize clusters of equal orientation. In our simulations of soft disks, we also have noticed empirically that the mean $|\psi_6|$ is roughly $\langle|\psi_6|\rangle$ = 0.69 at the liquid branch of the liquid to hexatic phase transition (see SI), and we use this parameter as an additional indicator for the transition. that we refer to $|\psi_6|$ as the average over the whole system throughout the paper. Only the colored snapshots visualize per particle $|\psi_6|$ values. For computing g(r) and $\psi_6$ we employ the MD analysis program FREUD[27].

Molecular Dynamics simulations of soft disks

We have performed Molecular Dynamics simulations of soft disks using the model of Kapfer and Krauth[17] with the HOOMD Molecular Dynamics package and a Langevin integrator[28]:

Eq. (4)

$$V(r) = \left(\frac{\sigma}{r}\right)^n$$

In this coarse, phenomenological model for the bulk behavior of skyrmions, σ roughly corresponds to the mean skyrmion distance, and n denotes the steepness of the potential. By running MD simulations at the experimentally determined skyrmion density, we were able to adjust the simulation potential so that the pair correlation for the simulated soft disks matches the pair correlation of the skyrmions.

In order not to overparameterize the mapping to the experimentally measured PCFs, we only adjusted *n* and set σ constant. Even though the position of the first peak of the PCFs is not necessarily identical with the σ of the simulation potential, this approximation turns out to be sufficiently accurate for the examined densities. Therefore, we set σ in the simulations to be the position of the first peak of the experimentally determined PCFs. We then run simulations for varying *n* in the range between 6 and 12 with 0.1 resolution. The matching of the simulated and experimental PCFs is determined as mean squared deviation measured up to the fourth maximum. This deviation shows a smooth dependence of *n* and a clear minimum which we took as best match to the experiment. The optimal *n* is typically around 10 (for T = 338 K) and somewhat lower for lower temperatures.

The determined density, σ and *n* allow us to run simulations mapped closely to the experiment and the estimated underlying experimental potential. For these mapped simulations we have determined the mean absolute value $\langle|\psi_6|\rangle$ which is somewhat dependent on the equilibration time of the simulations. If not mentioned otherwise, we have equilibrated our system for $10^6$ time steps before measurements were taken.

It should also be noted that our simulations employ a Langevin dynamics thermostat with a time step of $10^{-3}$ and could be further improved by including additional specific terms to account for gyrotropic dynamics [29–31]. We do not expect that our current static equilibrium results are affected because such terms do not contribute to the energy of the system and thus do not break detailed balance [31]. However, to analyze the dynamics of the system evolution in the future, such terms need to be considered.

**Acknowledgments:** The authors would like to thank Kurt Binder for insightful discussions and acknowledge funding from TopDyn, SFB TRR 146 and SFB TRR 173 Spin+X (project A01). The experimental part of the project was additionally funded by the Deutsche Forschungsgemeinschaft (DFG, German Research Foundation) project No. 403502522 (SPP 2137 Skyrmionics) and the EU (3D MAGIC ERC-2019-SyG 856538). We acknowledge financial support from the Horizon 2020 Framework Programme of the European Commission under FET-Open Grant No. 863155 (s-Nebula).

[*1] https://github.com/soft-matter/trackpy, doi: 10.5281/zenodo.12255

**Data availability**

The data that support the plots within this paper and other findings of this study are available from the corresponding author upon reasonable request.

# Skyrmion Lattice Phases in Thin Film Multilayer.

Jakub Zázvorka[1,2,+], Florian Dittrich[1,+], Yuqing Ge[1], Nico Kerber[1,3], Klaus Raab[1], Thomas Winkler[1], Kai Litzius[1,3,4], Martin Veis[2], Peter Virnau[1,3,*], Mathias Kläui[1,3,*].

1. Institut für Physik, Johannes Gutenberg-Universität Mainz, Mainz, Germany.
2. Charles University, Faculty of Mathematics and Physics, Institute of Physics, Ke Karlovu 5, CZ-121 16, Prague 2, Czech Republic.
3. Graduate School of Excellence Materials Science in Mainz, Mainz, Germany.
4. Max Planck Institute for Intelligent Systems, Stuttgart, Germany.

[+]*These authors contributed equally.*
[*] Corresponding authors: Klaeui@uni-mainz.de, Virnau@uni-mainz.de

## Supplementary Information

1. MOKE Loops

Magnetization hysteresis loops using the Kerr microscope were investigated with respect to an out-of-plane field. No in-plane field was applied. The measurements were performed to determine the range of the out-of-plane field which can be applied to stabilize domains and magnetic textures in the sample. The MOKE signal is evaluated as the overall grayscale contrast averaged over the whole picture frame. Figure S1 shows the gradual change of the hysteresis loops at different temperatures. Whereas at 300 K a sharp switching was observed and no skyrmions could be nucleated using field sweeping, at 325 K the sample shows a "hourglass-shaped" loop. In this configuration, using an out-of-plane field, a low density of skyrmions is nucleated together with elongated worm-like domains. At higher temperatures the loop tilts towards higher field and skyrmions are nucleated at those field values. However, the nucleation of skyrmion lattices was only achieved using in-plane field sweeping.

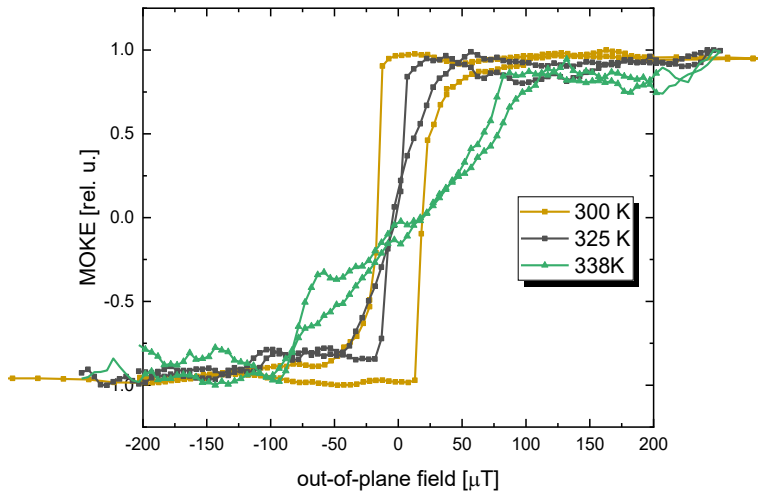

**Figure S1**. Magnetization hysteresis loops at various temperatures measured using the Kerr microscope. The hysteresis loops transform from rectangular sharp switching at temperature 300 K to a butterfly-shaped loop at temperatures around 330 K. At those temperatures the skyrmion lattices are nucleated and investigated.

2. Magnetization with Temperature

The temperature dependence of the magnetization was determined using the superconducting quantum interference device (SQUID). Measurements were performed to investigate the change of magnetization in the range of temperatures used in the study. The substrate contribution was obtained by measuring the magnetization loops and subtracting the diamagnetic background. Magnetization dependence was fitted using the Bloch's law. The Curie temperature was determined at $T_C \approx 448\ K$. This indicates that the temperature range used for the skyrmion lattice ordering is far enough from the critical temperature.

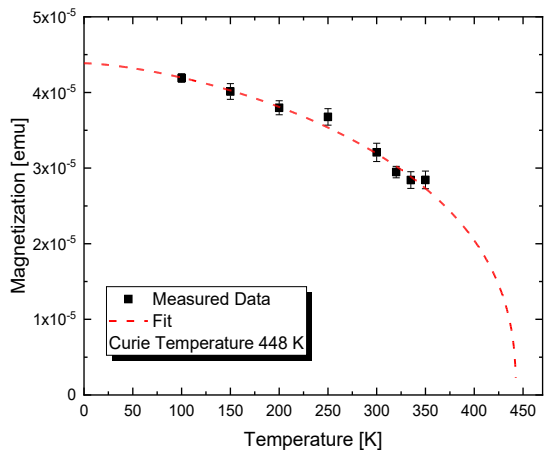

**Figure S2**. Temperature dependence of the magnetization of the studied material.

3. Skyrmion nucleation

The studied material exhibited perpendicular magnetic anisotropy. Using out-of-plane field sweeping, only stripe domains and a low density of skyrmions are present in the sample. Upon saturation of the sample using an in-plane field (not depending on the field orientation in the hard plane) and after discontinuation of the in-plane field, a high density of skyrmions is nucleated in the sample. The density and the mean radius of the skyrmions is controlled by simultaneous application of the out-of-plane field additional to the nucleation in-plane field. Depending on the interplay between the out-of-plane field and the material magnetic anisotropy, the nucleated skyrmions expand to worm domains or shrink in their size down to a complete annihilation. The magnetic state before and after the application of the in-plane field sweeping is shown in Figure S3. As it has been previously shown that skyrmion nucleation occurs in the time scale of nanoseconds and the time needed for the in-plane field to be fully switched in our setup is in the range of milliseconds, we cannot resolve the skyrmion nucleation by itself. However, we focus on the formation and effects of the skyrmion lattice which lie within the resolution of our setup. To evaluate the skyrmion lattice, videos of the domain state have been taken using the Kerr microscope. The starting point was after the application of the desired out-of-plane field and when the sample was fully saturated with the in-plane field. Afterwards, videos with the duration of one minute with the frame resolution of 16 fps (62.5 ms) have been taken for various out-of-plane field settings and several temperatures.

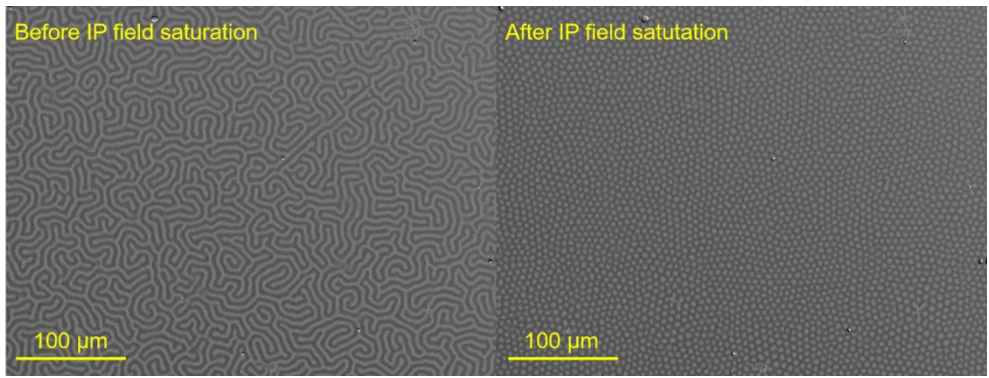

**Figure S3**. Effect of in-plane sweeping on the skyrmion nucleation at temperature 335 K and constant out-of-plane field of 20 µT. On the left is the domain state after application of the out-of-plane field. No skyrmions are present in this state. After in-plane (IP) field application and discontinuation, a high density of skyrmion is nucleated in the sample, as seen on the right.

4. Matching of experiment and simulation

The matching of simulations to experiments was outlined in the Methods Section and here we address more details.

To match a certain experimental measurement the positional data of the measured skyrmions are evaluated first. The pair correlation function is calculated averaging over the last 30s of the measurement video. The position of the (first) maximum of the so obtained g(r) sets the σ of our simulation potential (Eq. 4). The skyrmion density is also calculated by averaging over the last 30s of the measurement video. For each frame the number of detected skyrmions is divided by the detection area (given by the minimum and maximum of detected positions in x and y direction) and then averaged. This way the size of the unit cell of the simulation is given by the square root of the density. In different simulations with different exponents *n* in the potential, the deviation of the simulated g(r) from the experimental g(r) is calculated as described in the Methods Section. A plot of this deviation in dependence of *n* can be seen in figure S4a. It shows a clear minimum, from which *n* was determined. The simulated g(r) for this best match and the experimental one can be seen in figure S4b.

In general, this matching method works better for lower densities and higher out-of-plane fields. This is likely because the assumption that the position of the first peak roughly equals σ is more exact for these densities.

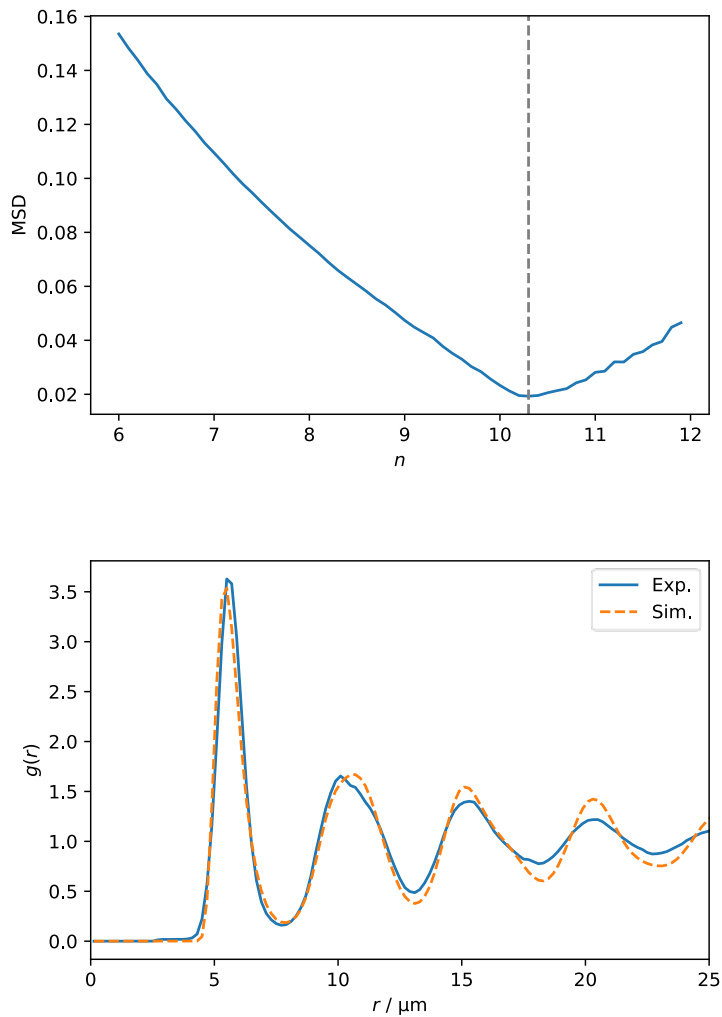

**Figure S4**. Matching theoretical simulations to experimental results. a) The mean squared deviation of the simulated pair correlation function and the experimental one, measured up to the 4th maximum in dependence of *n* (T = 338 K and 20 µT). The minimum is marked by the dashed line at *n* = 10.3. b) Experimental and matched pair correlation function (for T = 338 K and 20 µT), simulated with *n* = 10.3 as determined in figure S4a. The experimental pair correlation function is less pronounced in the higher order peaks, likely due to interfaces. Note that the matching works better for less dense systems. The experimental pair correlation function shows a small discontinuity at around 3 µm which corresponds to the cutoff of our tracking software. The slight increase for r>3 µm is likely due to defects.

5. Equilibration towards the hexatic phase

In order to better understand the equilibration process of the experimental data, we simulated the experimental data set at T = 338 K and 20 µT with corresponding mapped potential. As we showed before in Figure 1 the formation of the skyrmion lattice is taking place rather quickly (about 30s) which is indicated by the measurements of $\langle|\psi_6|\rangle$ and the pair correlation function. Measurements of $\langle|\psi_6|\rangle$ however continue to grow slowly, indicating that equilibrium is not reached yet. Figure 4 suggests, that the 60s of experimental measurements roughly correspond to something between $10^4$ to $10^5$ simulation steps for T = 338 K and 20 µT. Figure S5 shows the evolution of $\langle|\psi_6|\rangle$ for the same simulation in dependence of simulation steps. It shows that the equilibration towards the hexatic phase is still not reached for $10^6$ simulation steps and continues even further. This is in good agreement with the corresponding evolution of the spatial correlation functions shown in figure 4. Between $10^4$ to $10^5$ simulation steps $\langle|\psi_6|\rangle$ for the simulation is below the experimental $\langle|\psi_6|\rangle$. This might indicate that the mapped potential is a bit too soft due to the above-mentioned problems of finding the correct σ. In conclusion one can estimate, that a fully equilibrated skyrmion lattice at this experimental setup is deeper in the hexatic phase than the measured $\langle|\psi_6|\rangle$ and spatial correlation functions do indicate for now.

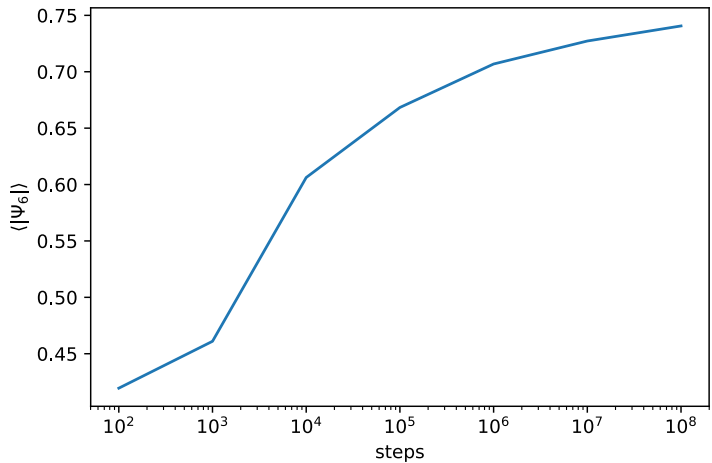

**Figure S5**. Time evolution of the mean absolute value $\langle|\psi_6|\rangle$ for a simulation matching the experimental 338 K, 20 µT system. Simulation took place in a quadratic box containing 40000 particles.

6. Analysis of $\langle|\psi_6|\rangle$ for soft disks

$\langle|\psi_6|\rangle$ for soft disks depends on the potential (i.e. *n*) and weakly depends on the system size. It turns out to be mostly independent of these parameters around the phase transition, namely the transition from the liquid phase to the liquid-hexatic-coexistence phase. This phase transition is the relevant one for us to mark the transition to the hexatic phase. Figure S6 visualizes the properties of $\langle|\psi_6|\rangle$ described above and shows that $\langle|\psi_6|\rangle$ is around 0.69 when the phase transition to the hexatic phase starts[1].

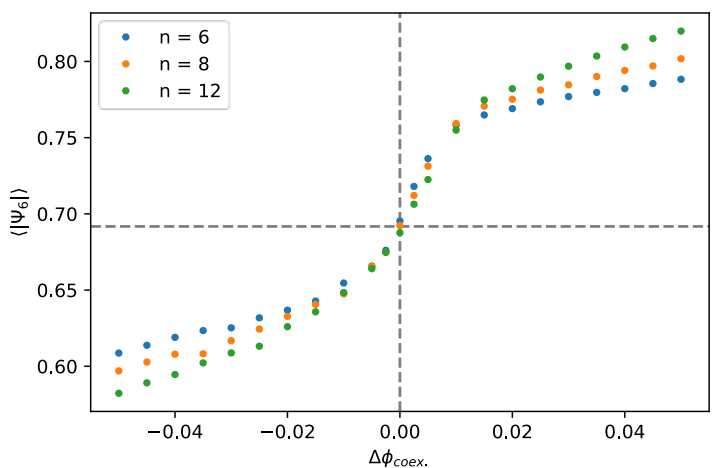

**Figure S6**. Shows that the mean absolute value $\langle|\psi_6|\rangle$ does not differ at the transition point and closely below for different potentials (with n= 6, 8 and 12). Same applies for different system sizes. Above points are the results of individual simulations of quadratic boxes containing 40000

particles each. X-Axis is linearly shifted to match the liquid-to-liquid-hexatic-coexistence phase transition points as determined in Ref.1. (dashed vertical line). The points in [-0.005...0.005] had $10^8$ steps equilibration time and were measured over the course of $10^9$ steps. All other points had $5 \cdot 10^6$ steps equilibration time and were measured over the course of $10^8$ steps.